\documentstyle[twocolumn,twoside]{article}
\newcommand{\DD}{{\cal D \hspace*{-2ex} D}}
\newcommand{\dee}{\mbox{$D \hspace*{-1.7ex} \raisebox{0.25ex}{$-$}$}}
\newcommand{\FF}{{\cal F \hspace*{-2ex} F}}
\newcommand{\plaision}{\, \rule[1.4ex]{1.4ex}{0.15ex} \hspace*{-1.4ex}
            \rule[0.17ex]{0.25ex}{1.3ex} \hspace*{-0.3ex} \Box}
\newcommand{\RR}{{\cal R \hspace*{-2.1ex} R}}
\newcommand{\smallvec}[1]{\vec{\scriptstyle \rm #1}}
\newcommand{\VV}{{\cal V \hspace*{-1.5ex} V}}
\newcommand{\WW}{{\cal W \hspace*{-2.6ex} W}}
\title{\Large \bf Five-Dimensional Tangent Vectors in Space-Time \\
\large \bf V. Generalization of Covariant Derivative}
\author{Alexander Krasulin \\ \it Institute for Nuclear Research of the
Russian Academy of Sciences \\ \it 60th October Anniversary Prospect, 7a,
117312 Moscow, Russia}
\date{\normalsize \bf Abstract \\ \mbox{ } \\ \begin{minipage}{400pt}
\normalsize In this part of the series I discuss the five-vector
generalizations of affine connection and gauge fields. I also give definition
to the exterior derivative of nonscalar-valued five-vector forms and consider
the five-vector analogs of the field strength tensor. In conclusion I discuss
the nonspacetime analogs of five-vectors. \end{minipage} }
\begin{document}

\maketitle

\begin{flushleft}
A. \it Five-vector affine connection
\end{flushleft}
As is known, the notion of the gradient for four-vector fields cannot be
formulated invariantly without introducing an additional structure on the
space-time manifold, which is called the affine connection. Formally, the
latter can be viewed as a map
\begin{equation}
\nabla : \; \DD \times \DD \rightarrow \DD,
\end{equation}
where $\DD$ is the set of all four-vector fields (derivations). The image
of a pair of fields $\bf (U,V)$ is denoted as $\bf \nabla_{U} V$ and is
called the covariant derivative of field $\bf V$ in the direction of field
$\bf U$. By definition, $\nabla$ satisfies the following three requirements:
\begin{flushright}
\hspace*{4ex} \hfill $\nabla_{(f{\bf U} + g{\bf V})} {\bf W} = f \cdot
\nabla_{\bf U} {\bf W} + g \cdot \nabla_{\bf V}{\bf W}$, \hfill (2a) \\
\hspace*{4ex} \hfill $\nabla_{\bf U} {\bf (V+W)} = \nabla_{\bf U} {\bf V}
+ \nabla_{\bf U} {\bf W}$, \rule{0ex}{3ex} \hspace{5ex} \hfill (2b) \\
\hspace*{4ex} \hfill $\nabla_{\bf U} (f{\bf V}) =
\partial_{\bf U} f \cdot {\bf V} + f \cdot \nabla_{\bf U}
{\bf V}$ \rule{0ex}{3ex} \hspace{5.5ex} \hfill (2c)
\end{flushright} \setcounter{equation}{2}
for any scalar functions $f$ and $g$ and any four-vector fields $\bf U, V$,
and $\bf W$. There exists a regular way in which the notion of the covariant
derivative can be extended to the fields of all other four-tensors. One
first defines it for an arbitrary 1-form field $\widetilde{\bf S}$:
$\bf \nabla_{U} \widetilde{S}$ is such that
\begin{equation} \bf
<\nabla_{U} \widetilde{S}, V> \; = \, \partial_{U} \! <\widetilde{S}, V>
 - <\widetilde{S}, \nabla_{U} V>
\end{equation}
for any four-vector field $\bf V$. The covariant derivatives of all other
four-tensor fields can then be defined by induction according to the formula
\begin{equation} \bf
\nabla_{U} (M \otimes N) = \nabla_{U} M \otimes N + M \otimes \nabla_{U} N,
\end{equation}
where $\bf M$ and $\bf N$ are any two four-tensor fields. For an arbitrary
scalar function $f$ one takes that
\begin{equation}
\nabla_{\bf U} f \equiv {\bf U} f = \partial_{\bf U} f.
\end{equation}

The meaning of all these formal definitions becomes apparent if one
interprets the value of $\bf \nabla_{U} V$ at each space-time point as
a derivative of field $\bf V$ along a parametrized curve whose tangent
four-vector is $\bf U$, calculated by using certain rules of parallel
transport. At an arbitrary point $P$, this latter derivative is constructed
by the following obvious procedure: $(i)$ Take $\bf V$ at $\lambda =
\lambda(P) + \Delta \lambda$. $(ii)$ Parallel transport it back to $P$.
$(iii)$ Calculate how much it differs from $\bf V$ there. $(iv)$ Divide by
$\Delta \lambda$ and take the limit $\Delta \lambda \rightarrow 0$. Any
derivative calculated this way satisfies requirement (2c) and is such that
\begin{equation}
\nabla_{(f{\bf U})} {\bf V} = f \cdot \nabla_{\bf U} {\bf V}
\end{equation}
for any $f$, $\bf U$, and $\bf V$, which is nothing but requirement (2a)
at $g = 0$. To satisfy requirement (2a) completely, $\nabla$ should also
have the property
\begin{equation}
\nabla_{({\bf U} + {\bf V})} {\bf W} = \nabla_{\bf U} {\bf W}
+ \nabla_{\bf V} {\bf W}
\end{equation}
for any $\bf U$, $\bf V$, and $\bf W$, which can be interpreted in the
following way: the parallel transport of four-vectors along any finite
continuous curve within any given region of space-time is completely
determined by the rules of parallel transport along the coordinate lines
of an arbitrary coordinate system (or systems) that covers the above region
completely. Requirement (2b) is equivalent to saying that parallel transport
is a linear operation. Equation (3) means that parallel transport conserves
the contraction of a four-vector and a four-vector 1-form. Equation (4)
means that the rules of parallel transport for different four-tensors are
correlated with one another in such a way that the tensor product of any two
four-tensors $\bf M$ and $\bf N$ is transported into the tensor product of
the transported $\bf M$ and transported $\bf N$. Equation (5) also becomes
obvious.

In part I the covariant derivative has been introduced for five-vector
fields. This is equivalent to introducing a map
\begin{equation}
\nabla : \; \DD \times \FF \rightarrow \FF.
\end{equation}
Considering the way five-vectors are related to four-vectors, one can
regard the structure defined on space-time by this map as an extension
of the structure defined on it by map (1). The subsequent replacement of
$\bf \nabla_{U}$ with the operator $\bf \nabla_{u} = \nabla_{(u^{\cal Z})}$
is equivalent to replacing map (8) with a map
\begin{equation}
\nabla : \; \FF_{\cal \! Z} \times \FF \rightarrow \FF,
\end{equation}
which is more a formality, since $\DD$ is isomorphic to $\FF_{\cal \! Z}$.
It now seems natural to make one more step in generalizing the concept of
affine connection to five-vectors and consider a map
\begin{equation}
\plaision : \; \FF \times \FF \rightarrow \FF,
\end{equation}
which will be called the {\em five-vector affine connection}. The image of a
pair of fields $\bf (u,v)$ with respect to $\plaision$ will be denoted as
$\bf \plaision_{u} v$ and will be called the {\em five-vector covariant
derivative} of field $\bf v$ in the direction of field $\bf u$. To give
$\plaision$ a formal definition, one should formulate certain requirements,
similar to requirements (2) for $\nabla$, which should be satisfied by
$\plaision$. The analogs of equations (2a) and (2b) are quite obvious:
\begin{flushright}
\hspace*{4ex} \hfill $\plaision_{(f{\bf u}+g{\bf v})}{\bf w} = f \cdot
\plaision_{\bf u}{\bf w} + g \cdot \plaision_{\bf v}{\bf w}$, \hfill (11a) \\
\hspace*{4ex} \hfill $\plaision_{\bf u} {\bf (v+w)} = \plaision_{\bf u}
{\bf v} + \plaision_{\bf u}{\bf w}$,\rule{0ex}{3ex} \hspace{5ex} \hfill (11b)
\end{flushright} \setcounter{equation}{11}
for any scalar functions $f$ and $g$ and any five-vector fields $\bf u$,
$\bf v$, and $\bf w$. To make a rational choice of the analog of requirement
(2c), let us first formulate explicitly the condition that the structure
defined on space-time by $\plaision$ is an extension of the structure defined
on it by $\nabla$. The latter statement apparently means that the restriction
of $\plaision$ to $\FF_{\cal \! Z} \times \FF$ should coincide with map
(9), which in its turn means that
\begin{equation}
\plaision_{(\bf u^{\cal Z})} = \nabla_{\bf u}
\end{equation}
for any five-vector field $\bf u$. Together with requirement (11a), the
latter equation yields
\begin{equation}
\plaision_{\bf u} = \nabla_{\bf u} + \varsigma \lambda_{\bf u} \,
\plaision_{\bf 1}.
\end{equation}
Since the five-vectors from $\cal E$ do not correspond to any direction
in space-time, let us assume that $\plaision_{\bf 1}$ is a {\em purely
algebraic} operator, so
\begin{equation}
\plaision_{\bf 1} (f {\bf v}) = f \cdot \plaision_{\bf 1} {\bf v}
\end{equation}
for any five-vector field $\bf v$ and any scalar function $f$. From the
latter equation and formula (13) one obtains the relation
\begin{flushright}
\hspace*{4ex} \hfill $\plaision_{\bf u}(f{\bf v}) = \partial_{\bf u}
f \cdot {\bf v} + f \cdot \plaision_{\bf u} {\bf v}$, \hfill (11c)
\end{flushright}
which is the desired analog of requirement (2c) for $\plaision$.

Let us now define the action of $\plaision$ on scalar functions. Considering
what has been said above, it seems reasonable to think that the action of
$\plaision_{\bf u}$ on an arbitrary scalar function $f$ should produce a
sum of the derivative $\partial_{\bf u} f$ and a term of the form
$a \lambda_{\bf u} f$, where $a$ is a constant. One should now notice that if
one adds to $\plaision$ a term proportional to $\lambda_{\bf u} {\bf 1}$, one
will obtain an operator that will still satisfy requirements (11), but whose
action on scalar functions will be different. In particular, one can select
this additional term in such a way that the action of the resulting operator
on $f$ would yield $\partial_{\bf u} f$. In the following, the notation
$\plaision_{\bf u}$ will refer to this particular choice of the five-vector
covariant derivative operator, and so
\begin{equation}
\plaision_{\bf u} f = \partial_{\bf u} f.
\end{equation}

As it follows from equations (11c) and (15), the action of $\plaision$ on
the product of two scalar functions and on the product of a scalar function
and a five-vector field obeys the Leibniz rule. Let us assume that the same
rule holds for the contraction and tensor product. This will enable us to
define the action of $\plaision$ on an arbitrary five-vector 1-form field
$\widetilde{\bf s}$ according to the formula:
\begin{equation} \bf
< \plaision_{u} \widetilde{s}, v> \; = \, \partial_{u} \!
< \widetilde{s}, v> - < \widetilde{s}, \plaision_{u} v>
\end{equation}
for any five-vector field $\bf v$, and, by induction, on the fields of all
other five-tensors according to the formula
\begin{equation} \bf
\plaision_{u} (m \otimes n) \, = \, \plaision_{u} m \otimes n \,
+ \, m \otimes \plaision_{u} n,
\end{equation}
where $\bf m$ and $\bf n$ are any two five-tensor fields.

There is one more constraint that should be imposed on $\plaision$, which
will enable one to define in a natural way the action of $\plaision$ on
four-vector fields. Namely, one should require that
\begin{equation}
{\bf v} \equiv {\bf w} \, ({\rm mod} \; R) \; \Longrightarrow \;
\plaision_{\bf u}{\bf v} \equiv \plaision_{\bf u}{\bf w} \, ({\rm mod} \; R),
\end{equation}
where $R$ is the equivalence relation considered in section 1 of part II.
The derivative $\plaision_{\bf u} {\bf V}$ of an arbitrary four-vector
field $\bf V$ can then be defined as the equivalence class with respect
to $R$ of all the fields of the form $\plaision_{\bf u}{\bf v}$, where
$\bf v \in V$.

Let us now consider the analogs of connection coefficients for $\plaision$.
For a given set of five-vector basis fields ${\bf e}_{A}$, it is natural to
define the latter according to the equation
\begin{equation}
\plaision_{A} {\bf e}_{B} = {\bf e}_{C} H^{C}_{\; BA},
\end{equation}
where $\plaision_{A} \equiv \plaision_{{\bf e}_{A}}$. The quantities
$H^{A}_{\; BC}$ will be called the {\em five-vector connection coefficients}.
If ${\bf e}_{A}$ is a regular basis, then from equation (12) it follows that
\begin{displaymath}
H^{A}_{\; B \mu} = G^{A}_{\; B \mu},
\end{displaymath}
where $G^{A}_{\; B \mu}$ are the connection coefficients associated with
$\nabla$. Furthermore, from condition (18) it follows that in any standard
five-vector basis
\begin{equation}
H^{\alpha}_{\; 5B} = 0
\end{equation}
at all $\alpha$ and $B$. In the usual way one can obtain the transformation
formula for five-vector connection coefficients corresponding to the
transformation ${\bf e}'_{A} = {\bf e}_{B} L^{B}_{\; A}$:
\begin{displaymath}
H'^{A}_{\; BC} = (L^{-1})^{A}_{\, D} \, H^{D}_{\; EF} \, L^{E}_{\, B} \,
L^{F}_{\, C} + (L^{-1})^{A}_{\, D} \, (\partial_{F} L^{D}_{\, B}) \,
L^{F}_{\, C}.
\end{displaymath}
If both bases are standard, one has
\begin{equation}
H'^{A}_{\; B5} = (L^{-1})^{A}_{\; D} \, H^{D}_{\; E5} \, L^{E}_{\; B} \,
L^{5}_{\; 5},
\end{equation}
so the coefficients $H^{A}_{\; B5}$ transform as components of a five-tensor
and cannot be nullified at a given space-time point by an appropriate choice
of the five-vector basis.

\vspace{3ex} \begin{flushleft}
B. \it Interpretation of the five-vector covariant \\
\hspace*{2ex} derivarive
\end{flushleft}
In the previous section we have introduced the five-vector covariant
derivative and have discussed its basic properties in a formal way. There
now arises a natural question: what is the meaning of this derivative?

The very way the operator $\plaision$ has been introduced in the previous
section suggests that the five-vector covariant derivative of an arbitrary
five-vector or five-tensor field $\cal S$ should be regarded as a
generalization of the ordinary covariant derivative of $\cal S$. From
equations (12)--(14) we see that technically this generalization consists
in that to the four quantities $\nabla_{\alpha} \cal S$, which characterize
the variation of $\cal S$ in the vicinity of a given space-time point and
which can be interpreted as derivatives of $\cal S$ in the direction of the
basis five-vectors ${\bf e}_{\alpha}$, calculated by using certain rules of
parallel transport, one adds a fifth quantity, $\plaision_{5} \cal S$, which
depends only on the value of $\cal S$ at the considered point and which, by
itself, cannot be interpreted as a derivative of $\cal S$ in any direction.
These observations lead one to the simplest and the most obvious
interpretation of the five-vector covariant derivative where the latter
is broken up into a differential and a local part, each of which is then
interpreted independently. Since the interpretation of the differential
part of $\plaision$, which has all the properties of an ordinary covariant
derivative, is quite obvious, the question about the meaning of the
five-vector covariant derivative of $\cal S$ is reduced to the question
about the meaning of the quantity $\plaision_{5} \cal S$: what does it
characterize and what role is played by the length of the fifth basis vector?

Having in mind the possible application of $\plaision$ in physics, for
example, in field theory, and assuming that this application consists in
replacing the ordinary derivative with the operator $\plaision$ in the
equations of motion and in the expressions for the relevant physical
quantities, one can answer the above questions in the following, somewhat
formalistic way: the fifth component of the five-vector covariant derivative
is just a quantity that appears in the equations of motion and in other
physical formulae and which, one may think, expresses some new local
properties of space-time; since these formulae involve gradients (exterior
derivatives) and not derivatives in some definite directions, in the
case of the five-vector covariant derivative, too, one will apparently deal
with the operator $\plaision \equiv \widetilde{\bf o}^{A} \, \plaision_{A}$,
which does not depend on the choice of the five-vector basis and, in
particular, on the length of the fifth basis vector.

It should be noted that within this approach to the interpretation of
$\plaision$, the parallel transport of vectors is determined only by the
differetial part of $\plaision$, i.e.\ by the operator $\nabla$, and
consequently has all the usual properties, including the one expressed
by equation (7).

There exists another way of interpreting the five-vector covariant derivative
where the latter as a whole is regarded as a derivative in some direction,
calculated by using certain rules of parallel transport. Since the properties
of $\plaision$ differ from those of the ordinary covariant derivative, it is
{\em a priori} obvious that the properties of the parallel transport
associated with $\plaision$ as a whole should in some way be different from
the usual ones. To gain a better understanding of this issue, let us begin
by considering a general situation where one is given certain rules of
parallel transport, by using which one can evaluate the derivatives of
fields, and then determine what properties this transport should have in
order that the operator of the derivative it defines could be identified
with the operator $\plaision$. It seems reasonable to suppose that if such a
transport has any physical meaning, then the rules according to which it is
performed can in principle be found by analyzing the motion of particles and
light rays. In view of this, let us assume that initially one knows only
the rules of parallel transport along the timelike and null curves in the
positive direction of time. For simplicity, let us consider the transport
of four-vectors, though the analysis presented below will be essentially
valid for the case of five-vector transport as well, and even for the case
where the transported objects are some kind of abstract vectors or tensors
that are not directly related to space-time.

Already at this stage one can say whether or not the consired transport has
three important properties: linearity, the property of conserving the inner
product $g$, and torsion. In the following it will always be assume that
parallel transport is linear and conserves $g$. In that case the absence
or existence of torsion can be established by observing whether or not the
considered transport coincides with the transport defined by space-time
metric (i.e.\ by the torsion-free $g$-conserving ordinary covariant
derivative, which is uniquely fixed by the metric). In the following I will
always allow for arbitrary four-vector torsion (the notion of torsion
for five-vectors will be discussed in the next part of the series).

Knowing the considered rules of parallel transport, one can evaluate the
derivative of any sufficiently smooth four-vector field along any timelike or
null parametrized curve at any point $P$ in space-time. Since one knows the
rules of transport only in the positive direction of time, such a derivative,
strictly speaking, should be calculated not according to the procedure
described in the beginning of the previous section, but according to the
procedure where the value of the field at $P$ is compared with its value at
the point $\lambda (P) - \Delta \lambda$, and not at the point $\lambda (P)
+ \Delta \lambda$. The derivative in question will, naturally, depend on the
direction of the curve at the considered point and on the rate with which its
parameter changes. This dependence will be such that the derivative can be
considered a function of the four-vector tangent to the curve. This function
will be homogeneous, though not necessarily linear. In the general case, let
us denote the operator of this derivative as $D({\bf U})$, where $\bf U$
is the mentioned tangent four-vector. So far $D$ has been defined only for
the timelike and null four-vectors directed towards the future, the set of
which will be denoted as $V^{+}_{4}$.

As it has been said in section A, any derivative defined in such a way has
the properties expressed by equations (2c) and (6), where the operator
$\nabla_{\bf U}$ should now be replaced with the operator $D({\bf U})$.
Owing to the linearity of the considered transport, this derivative will
also have the property expressed by equation (2b), where one should make
a similar replacement. To determine the dependence of $D({\bf U})$ on $\bf U$
more precisely, let us consider at each space-time point the quantity
\begin{equation}
\Lambda({\bf U,V}) \equiv D({\bf U+V}) - D({\bf U}) - D({\bf V}).
\end{equation}
By virtue of the analog of property (2c) for $D({\bf U})$ and owing to the
linear dependence of the derivative $\partial_{\bf U}$ on its argument,
$\Lambda({\bf U,V})$ is a {\em local} operator:
\begin{displaymath}
\Lambda({\bf U,V}) \, (f {\bf W}) = f \cdot \Lambda({\bf U,V}) \, {\bf W}
\end{displaymath}
for any scalar function $f$ and any four-vector field $\bf W$. It is evident
that operator (22) characterizes the relation between the derivatives $D$
in different timelike and null directions. Usually, it is taken that the
rules of parallel transport along different curves are correlated with one
another in such a way that
\begin{equation}
\Lambda({\bf U,V}) = 0
\end{equation}
for all $\bf U$ and $\bf V$ from $V^{+}_{4}$. The latter equation is
evidently nothing but property (7) of the ordinary covariant derivative
and means that $D$ depends on its argument linearly. Equation (7) is
satisfied necessarily if four-vector torsion is zero.

If equation (23) holds at every point, $D({\bf U})$ is an ordinary covariant
derivative, and instead of $D({\bf U})$ I will then write $\nabla_{\bf U}$.
In this case one can define in a natural way the operator $\nabla_{\bf U}$
for all other $\bf U$ from $V_{4}$ by postulating that parallel transport
is reversible:
\begin{equation}
\nabla_{\bf (-U)} = - \nabla_{\bf U},
\end{equation}
and that equation (7) holds for timelike and null four-vectors with
{\em any} time orientation. Since any four-vector can be presented as a
sum or difference of two four-vectors from $V^{+}_{4}$, properties (7)
and (24) fix the operator $\nabla_{\bf U}$ for any $\bf U$.

Let us now consider the case where equation (23) is {\em not} obeyed.
Obviously, there exist an infinite number of ways of how this can be so,
and though in all cases the operator $\Lambda({\bf U,V})$ should be a
symmetric function of its arguments and have properties like
\begin{displaymath}
\Lambda(k {\bf U}, k {\bf V}) = k \cdot \Lambda({\bf U,V}) \; \; \; (k > 0)
\end{displaymath}
and
\begin{displaymath}
\Lambda(k {\bf U}, l {\bf U}) = 0 \; \; \; (k,l > 0),
\end{displaymath}
which evidently follow from definition (22), these constraints limit the
possible form of $\Lambda({\bf U,V})$ very little. To obtain a derivative
that could be identified with $\plaision$, let us consider one of the
simplest cases where for all $\bf U$ and $\bf V$ from $V^{+}_{4}$ the
operator $\Lambda({\bf U,V})$ is {\em proportional to the same local
operator}. It is evident that the normalization of the latter can be chosen
arbitrarily. Let us suppose that we have fixed it somehow, and let us
denote the corresponding operator as $\Delta$. The above assertion about
$\Lambda({\bf U,V})$ can then be expressed in the following way:
\begin{equation}
\Lambda({\bf U,V}) = \phi({\bf U,V}) \cdot \Delta,
\end{equation}
where $\phi({\bf U,V})$ is some real-valued symmetric function of two
four-vectors from $V^{+}_{4}$, which has several other properties that
follow from the properties of $\Lambda({\bf U,V})$ mentioned above.
Naturally, at different points in space-time the function $\phi({\bf U,V})$
and the operator $\Delta$ itself may be different.

Let us select a basis in the tangent space of four-vectors at the considered
point in such a way that all ${\bf E}_{\alpha} \in V^{+}_{4}$. Let us choose
an arbitrary four-vector ${\bf U} = U^{\alpha}{\bf E}_{\alpha}$ from
$V^{+}_{4}$ and consider for it the quantity
\begin{displaymath}
D({\bf U}) - U^{\alpha} D({\bf E}_{\alpha}).
\end{displaymath}
By using equation (25) it is not difficult to prove that for any such
$\bf U$ the above quantity is proportional to $\Delta$. Denoting the
proportionality factor as $\varrho({\bf U})$, one can write:
\begin{equation}
D({\bf U}) = U^{\alpha} \Delta_{\alpha} + \varrho({\bf U}) \cdot \Delta,
\end{equation}
where $\Delta_{\alpha} \equiv D({\bf E}_{\alpha})$. Consequently, the set of
operators $D({\bf U})$ for all ${\bf U} \in V^{+}_{4}$ at the considered
point is a subset of some five-dimensional real vector space.

  From equation (26) it follows that $\varrho({\bf U})$ is a homogeneous
function of $\bf U$, since this is true of the first term in the right-hand
side of (26) and of the operator $D({\bf U})$ itself. However, the
dependence of $\varrho({\bf U})$ on $\bf U$ cannot be linear, since in
that case the operator $D({\bf U})$ would also be linear in $\bf U$, and
$\varrho({\bf U})$ would be identically zero. Therefore, in equation (26)
the operator $D({\bf U})$ is presented as a sum of a term linear in $\bf U$
and of a term proportional to $\Delta$ and depending on $\bf U$ nonlinearly.
It is evident that such a decomposition is not unique. Indeed, one can write
that
\begin{displaymath}
D({\bf U}) = U^{\alpha} \Delta'_{\alpha} + \varrho'({\bf U}) \cdot \Delta,
\end{displaymath}
where
\begin{flushright}
\hspace*{4ex} \hfill $\Delta'_{\alpha} = \Delta_{\alpha} + X_{\alpha}
\cdot \Delta$ \hspace*{3.5ex} \rule{0ex}{0ex} \hfill {\rm (27a)} \\
\hspace*{4ex} \hfill $\varrho'({\bf U}) = \varrho({\bf U}) - U^{\alpha}
X_{\alpha},$ \rule{0ex}{3ex} \hfill {\rm (27b)}
\end{flushright} \setcounter{equation}{27}
and $X_{\alpha}$ are arbitrary constants. The function $\varrho'$
has the same formal properties as $\varrho$. Moreover, if instead of
the basis ${\bf E}_{\alpha}$ one selects some other four-vector basis,
${\bf E}'_{\alpha}$, also made only of vectors from $V^{+}_{4}$, then,
as one can easily prove, the function $\varrho'$ corresponding to it
will be related to $\varrho$ by a transformation of the form (27b). From
equation (26) it follows that the function $\varrho$ corresponding to a
given basis ${\bf E}_{\alpha}$ satisfies the condition
\begin{displaymath}
\varrho({\bf E}_{\alpha}) = 0 \; \mbox{ for all } \alpha.
\end{displaymath}

It is easy to see that the function $\phi$ introduced above is expressed
in terms of $\varrho$ as follows:
\begin{equation}
\phi({\bf U,V}) = \varrho({\bf U+V}) - \varrho({\bf U}) - \varrho({\bf V}),
\end{equation}
and is invariant, as it should be, under transformation (27b). From formula
(28) and the homogeneity of $\varrho$ follow the symmetry of $\phi$ and all
its other properties that can be derived from definition (22). Therefore,
equation (26) does not impose any constraints on $\varrho$ except for
homogeneity. Other than that this function can be absolutely arbitrary.

To obtain a derivative that can be identified with $\plaision$, let us
impose one more constraint on the considered transport of four-vectors,
which can be substantiated by the following arguments.

One should observe that in its structure, the expression in the right-hand
side of formula (26) is a contraction of the five numbers $U^{0}$, $U^{1}$,
$U^{2}$, $U^{3}$, $\varrho({\bf U})$ with the five operators $\Delta_{0}$,
$\Delta_{1}$, $\Delta_{2}$, $\Delta_{3}$, $\Delta$. The first four numbers
characterize only the direction and parametrization of the curve along which
the derivative is calculated, and do not depend in any way on the rules of
parallel transport, all the information about which (in this part of
$D({\bf U})$) is contained in the quantities $\Delta_{\alpha}$. Considering
this, it seems natural to examine the case where $\varrho({\bf U})$, too, is
independent of the transport rules and is only a characteristic of the curve.
Though formally this assumption does not place any limitations on the form
of $\varrho({\bf U})$, in fact it means that at every point in space-time
the latter should be invariant under active Lorentz transformations in the
tangent space of four-vectors, since otherwise space-time would acquire a
local anisotropy of unknown origin, which physically is not a very appealing
idea. Since the decomposition of $D$ into a linear and a nonlinear part is
not unique, the above assertion about the invariance of $\varrho$ should be
formulated as follows: there exists such a choice of the constants
$X_{\alpha}$ in formula (27b) that the function $\varrho'$ obtained by
this transformation is invariant under the mentioned active transformations
in $V_{4}$. Since in addition to this, $\varrho'({\bf U})$ is a homogeneous
function of $\bf U$, it should simply be proportional to the length of
$\bf U$. The proportionality factor can be absorbed into the operator
$\Delta$, whose normalization up to this point has been arbitrary, so at
an appropriate choice of $X_{\alpha}$ one will have
\begin{equation}
D({\bf U}) = U^{\alpha} \Delta'_{\alpha} + \| {\bf U} \| \cdot \Delta,
\end{equation}
where $\| {\bf U} \| \equiv \sqrt {g({\bf U,U})}$ and the operators
$\Delta'_{\alpha}$ are given by formula (27a). It is easy to prove that
owing to the nonlinear dependence of $\| \bf U \|$ on $\bf U$, the function
$\varrho'({\bf U})$ can be proportional to $\| {\bf U} \|$ only at one
particular choice of the constants $X_{\alpha}$.

Up to now we have considered the operator $D$ a function of the tangent
four-vector. Since $V_{4}$ is isomorphic to $\cal Z$, one can just as well
consider $D$ a function of the five-vector from $\cal Z$ that corresponds
to the tangent four-vector. One can then repeat the analysis made in this
section and obtain the analogs of all the formulae presented above, where
instead of the four-vectors from $V^{+}_{4}$ there will now stand the
corresponding five-vectors from $\cal Z$, the set of which will be denoted
as ${\cal Z}^{+}$. In particular, the analogs of formulae (27) will have
the form
\begin{flushright}
\hspace*{4ex} \hfill $\Delta'_{\alpha} = \Delta_{\alpha} + X_{\alpha}
\cdot \Delta$ \hspace*{2ex} \rule{0ex}{0ex} \hfill {\rm (30a)} \\
\hspace*{4ex} \hfill $\varrho'({\bf u}) = \varrho({\bf u}) - u^{\alpha}
X_{\alpha},$ \rule{0ex}{3ex} \hfill {\rm (30b)}
\end{flushright} \setcounter{equation}{30}
where $\Delta_{\alpha} \equiv D({\bf e}_{\alpha})$, ${\bf e}_{\alpha}$ is
some basis in $\cal Z$ consisting only of five-vectors from ${\cal Z}^{+}$,
and $u^{\alpha}$ are the components of the arbitrary five-vector ${\bf u}
\in {\cal Z}^{+}$ in this basis. Formula (29) will acquire the form
\begin{equation}
D({\bf u}) = u^{\alpha} \Delta'_{\alpha} + \| {\bf u} \| \cdot \Delta,
\end{equation}
where $\| {\bf u} \| \equiv \sqrt{g({\bf u,u})}$. It should be noted that
the operator $\Delta$ can be found from the equation
\begin{displaymath}
\Lambda({\bf u,v}) = ( \| {\bf u+v} \| - \| {\bf u} \| - \| {\bf v} \| )
\cdot \Delta
\end{displaymath}
at any such ${\bf u}$ and ${\bf v}$ from ${\cal Z}^{+}$ that the expression
in the brackets in the right-hand side does not vanish. The operators
$\Delta'_{\alpha}$ can be found according to the formula
\begin{equation}
\Delta'_{\alpha} = D({\bf e}_{\alpha}) - \| {\bf e}_{\alpha} \| \cdot \Delta.
\end{equation}

In the general case, the operator $D({\bf u})$ is a nonlinear function of
$\bf u$. As is seen from formula (31), one can make the dependence of $D$
on its argument linear if instead of considering it a function of ${\bf u}
\in {\cal Z}^{+}$, one formally regards it as a function of a five-vector
whose $\cal Z$-component coincides with $\bf u$ and whose $\cal E$-component
is proportional to $\| {\bf u} \|$. Denoting this latter five-vector
as $\breve{\bf u}$, from dimensional considerations one finds that
$\breve{\bf u}^{\cal E}$ should equal $\| {\bf u} \| \cdot k {\bf n}$,
where $\bf n$ is the normalized five-vector from $\cal E$ introduced in
part II and $k$ can be any nonzero real number. Without any loss in
generality one can put $k=1$, considering this a part of the definition
of $\breve{\bf u}$. Thus, one will have
\begin{equation}
\breve{\bf u} = {\bf u} + \| {\bf u} \| \cdot {\bf n}.
\end{equation}
One can regard $\breve{\bf u}$ as a tangent vector of a new type. Considering
that it is a homogeneous function of the directional derivative operator, one
may call it a {\em homogeneous} tangent five-vector. It is easy to see that
$\breve{\bf u}$ satisfies the relation
\begin{equation}
h(\breve{\bf u},\breve{\bf u}) = (1 + {\rm sign} \xi )
\cdot g(\breve{\bf u},\breve{\bf u}),
\end{equation}
which can be used as a condition that fixes the $\cal E$-component of
$\breve{\bf u}$ up to a sign.

To distinguish the operator $D$ regarded as a function of a five-vector from
${\cal Z}^{+}$ from the same operator regarded as a function of a
homogeneous tangent five-vector, in the second case instead of $D$ I will
use the symbol $\dee$. Initially, the operator $\dee$ is defined only for
the five-vectors of the form (33) with ${\bf u} \in {\cal Z}^{+}$. One can
then define it for all other five-vectors by linearity, i.e.\ supposing that
for any ${\bf w} \in V_{5}$
\begin{displaymath}
\dee ({\bf w}) = w^{\alpha} \Delta'_{\alpha} + w^{5} \Delta,
\end{displaymath}
where $w^{A}$ are the components of $\bf w$ in the normalized regular basis
${\bf e}'_{A}$ whose first four elements coincide with the basis vectors
${\bf e}_{\alpha}$ from ${\cal Z}^{+}$ considered above. Then, for
$\Delta'_{\alpha}$ and $\Delta$ one will have
\begin{displaymath}
\Delta'_{\alpha} = \dee ({\bf e}'_{\alpha}) \; \; \mbox{ and } \; \;
\Delta = \dee ({\bf e}'_{5}).
\end{displaymath}
In a similar manner, from the definition of operators $\Delta_{\alpha}$ and
equation (32) one obtains
\begin{displaymath}
\Delta_{\alpha} = \dee ({\bf e}'_{\alpha} - X_{\alpha} {\bf e}'_{5}),
\end{displaymath}
where $X_{\alpha} = - \, \| {\bf e}'_{\alpha} \|$. In the light of these
relations, formulae (30) acquire a new meaning: formula (30a) expresses
the operator $\dee ({\bf e}'_{\alpha})$ in terms of operators
$\dee ({\bf e}''_{\alpha})$ and $\dee ({\bf e}''_{5})$, where
${\bf e}''_{\alpha} = {\bf e}'_{\alpha} - X_{\alpha} {\bf e}'_{5}$ and
${\bf e}''_{5} = {\bf e}'_{5}$, and equation (27b) is the transformation
formula for the fifth component of $\breve{\bf u}$ corresponding to the
transformation from the basis ${\bf e}''_{A}$ to the basis ${\bf e}'_{A}$.

Up to now we have been talking about the transport of four-vectors. In an
absolutely similar way we could have considered the transport of five-vectors
and would have obtained the derivative $D$ or $\dee$ defined for five-vector
fields. I have preferred to deal with four-vectors because the situation
with their parallel transport is simpler: this transport can be absolutely
arbitrary as long as it conserves the linear operations and the inner product
$g$. As for the transport of five-vectors, so far we have considered only one
particular case, examined in section 3 of part II, and have not yet discussed
what requirements---in addition to linearity, conservation of $g$, and
preservation of the equivalence relation $R$---this transport should satisfy
in the general case. This uncertainty is of no importance to our analysis
though, since here we are concerned only with the {\em relation} between the
derivatives of fields in different space-time directions.

Repeating the analysis performed above for the case of five-vector parallel
transport, one will obtain a derivative $\dee$ whose operator will have the
same formal properties as $\plaision$. Indeed, for five-vector fields $\dee$
will be a map from $\FF \times \FF$ to $\FF$, which, as has been said above,
will have properties (11b), (11c), and (15), and, owing to the linear
dependence of $\dee$ on its argument, also property (11a). If in addition
to this the transport of five-vectors conserves the contration and tensor
product and preserves the equivalence relation $R$, then $\dee$ will also
have properties (16), (17), and (18). The derivative $\plaision$ can
therefore be interpreted as a derivative in a certain direction, calculated
by using certain rules of parallel transport whose properties differ from
the usual ones. From now on, instead of $\dee$ I will write $\plaision$.

When deriving the properties of $\plaision$ from the properties of the
corresponding transport, it is convenient to use the following proposition:
if equation
\begin{displaymath}
\plaision_{\bf v} {\cal S} = 0,
\end{displaymath}
where $\cal S$ can be a field of arbitrary nature, holds for any homogeneous
tangent five-vector $\bf v$ with ${\bf v}^{\cal Z} \in {\cal Z}^{+}$, then
it also holds for {\em any} $\bf v$ from $V_{5}$. This theorem follows
evidently from the fact that $\plaision$ is a linear function of its argument
and the fact that one can select a basis in $V_{5}$ consisting only of
homogeneous tangent five-vectors with the $\cal Z$-component belonging to
${\cal Z}^{+}$. By using this theorem one can easily show that from the
conservation of $g$ by the considered transport of five-vectors follows
the equation
\begin{equation}
\plaision g = 0,
\end{equation}
in which $g$ is regarded as a five-tensor. From this equation follow
certain constraints on the five-vector connection coefficients
$H^{A}_{\; BC}$, which are similar to those constraints on the four-vector
connection coefficients that follow from the equation $\nabla g=0$.

In conclusion, let me say a few words about the definition of the transport
corresponding to $\plaision$ for the timelike and null curves directed
towards the past and for the spacelike curves. From everything that has been
said above it is apparent that the problem actually comes to selecting for
the mentioned curves the homogeneous tangent five-vector.

Ordinary parallel transport is reversible, which means that the derivative
corresponding to it changes its sign whenever the parameterization of the
curve is changed for the opposite ($\lambda \rightarrow - \lambda$). If the
same requirement is imposed on the transport corresponding to $\plaision$,
then the homogeneous tangent five-vector corresponding to a timelike or
null curve directed towards the past will be given by the formula
\begin{displaymath}
\breve{\bf u} = {\bf u} - \| {\bf u} \| \cdot {\bf n},
\end{displaymath}
and not by formula (33). For spacelike curves the $\cal E$-component of
$\breve{\bf u}$ will be zero altogether, since on the one hand, it should
change sign whenever the parameterization of the curve is reversed, and
on the other hand, it should remain the same since the tangent four-vectors
corresponding to these two parametrizations can be transformed one into
the other by a Lorentz transformation. The latter means that for spacelike
curves the transport associated with $\plaision$ will coincide with the
transport corresponding only to the differential part of $\plaision$.

This asymmetry in the definition of the homogeneous tangent vector between
the spacelike and non-spacelike curves and between the curves directed
towards the future and towards the past may be interpreted as an indication
to that in the case of parallel transport associated with $\plaision$ the
condition of reversibility should not be imposed. Instead, one can postulate
that for all types of curves the homogeneous tangent five-vector is given
by formula (33), where for the spacelike curves $\| {\bf u} \|$ means
$\sqrt{-g({\bf U,U})}$. The corresponding transport will differ from the
ordinary one by an additional rotation of the transported vectors, which
will not depend on the direction of the transport but only on the length
of the travelled path, and it is this additional twist that will produce
the ``local'' part of the operator $\plaision$.

\vspace{3ex} \begin{flushleft}
C. \it Five-vector covariant derivative for fields \\
\hspace*{2ex} of nonspacetime vectors and tensors
\end{flushleft}
The five-vector covariant derivative can also be defined for the fields whose
values are some kind of abstract vectors or tensors that have no direct
relation to the space-time manifold. In the following such vectors and
tensors will be referred to as {\em nonspacetime} vectors and tensors.

Let us consider a set $\VV$ of fields whose values are some $n$-dimensional
nonspacetime vectors, which will be denoted with small capital Roman letters
with an arrow: $\smallvec{A}, \smallvec{B}, \smallvec{C}$, etc. Defining an
ordinary covariant derivative for such fields is equivalent to fixing a map
\begin{displaymath}
\nabla : \; \DD \times \VV \rightarrow \VV,
\end{displaymath}
or an equivalent map
\begin{equation}
\nabla : \; \FF_{\cal \! Z} \times \VV \rightarrow \VV,
\end{equation}
which should satisfy three requirements similar to requirements (2) for
map (1). If $\smallvec{E}_{i}$ ($i = 1, \ldots, n$) is some set of basis
fields in $\VV$, then the corresponding connection coefficients, which are
called gauge fields in physics, are defined by the equation
\begin{equation}
\nabla_{\mu} \smallvec{E}_{i} = \smallvec{E}_{j} A^{j}_{\; i \mu}.
\end{equation}

In a similar manner one can formally define the five-vector covariant
derivative for the fields from $\VV$. This is equivalent to fixing a map
\begin{equation}
\plaision : \; \FF \times \VV \rightarrow \VV,
\end{equation}
which will be regarded as an extension of map (36), so in this case, too,
the operators $\plaision$ and $\nabla$ will be related as in equation (12).
In addition to this, map (38) should satisfy three requirements similar to
requirements (11) for five-vector fields, which I will not present here.

The connection coefficients corresponding to derivative (38), which I will
call {\em five-vector gauge fields}, are defined by the equation
\begin{equation}
\plaision_{A} \smallvec{E}_{i} =  \smallvec{E}_{j} B^{j}_{\; i A}.
\end{equation}
It is apparent that in any regular five-vector basis
\begin{displaymath}
B^{i}_{\; j \mu} = A^{i}_{\; j \mu}
\end{displaymath}
for any $i$, $j$, and $\mu$. In the usual manner one can obtain the formula
for transformation of five-vector gauge fields under the transformation
$\smallvec{E} \, '_{i} = \smallvec{E}_{j} L^{j}_{\; i}$ of the basis fields
in $\VV$:
\begin{displaymath}
B'^{\, i}_{\, \; jA} = (L^{-1})^{i}_{\; k} \, B^{k}_{\; lA} \, L^{l}_{\; j}
+ (L^{-1})^{i}_{\; k} \, \partial_{A} L^{k}_{\; j}.
\end{displaymath}
  From this formula it follows that in any standard five-vector basis
\begin{equation}
B'^{\, i}_{\, \; j 5}=(L^{-1})^{i}_{\; k} \, B^{k}_{\; l 5} \, L^{l}_{\; j},
\end{equation}
so the fields $B^{i}_{\; j5}$ transform as components of a tensor of
rank $(1,1)$ over $\VV$. This latter fact, together with the facts that
$B^{i}_{\; j5}$ are Lorentz scalars and that in the equations of motion
for matter fields they will appear at the place where the mass parameter
usually stands, may suggest that some of these new gauge fields can
effectively play the role of Higgs fields.

Let us now suppose that for the considered type of nonspacetime vectors
there exists a certain nondegenerate inner product $\theta$. If this inner
product is conserved by the parallel transport associated with $\plaision$,
then from considerations similar to those that have led to equation (35)
it follows that
\begin{displaymath}
\plaision_{A} \, \theta = 0,
\end{displaymath}
where $\theta$ is regarded as a tensor over $\VV$. This equation imposes
certain constraints on the five-vector gauge fields. For example, if the
inner product is Hermitian, these fields should be such that
\begin{equation}
\partial_{A} \theta_{ij} - \theta_{kj} (B^{\, k}_{\, \; iA})^{\ast}
- \theta_{ik} B^{\, k}_{\, \; jA} = 0,
\end{equation}
where $\scriptstyle \ast$ denotes complex conjugation. If at each point in
space-time the basis $\smallvec{E}_{i}$ is selected orthonormal, then from
the latter equation one obtains that
\begin{displaymath}
\theta_{kj} (B^{\, k}_{\, \; iA})^{\ast} + \theta_{ik} B^{\, k}_{\, \; jA}
= 0,
\end{displaymath}
which means that the quantities $B_{ijA} \equiv \theta_{ik} B^{\, k}_{\, \;
jA}$ are anti-Hermitian matrices with respect to the indices $i$ and $j$.

\vspace{3ex} \begin{flushleft}
D. \it Exterior derivative of nonscalar-valued forms
\end{flushleft}
In applications of exterior differential calculus one often has to deal
with forms whose values are not scalars but are some vector-like objects:
four-vectors or four-tensors; spinors; some kind of nonspacetime vectors
or tensors; or tensor products of such objects. To this list one can now
add five-vectors and five-tensors.

For obvious reasons, in the general case an arbitrary nonscalar-valued form
$\widetilde{\bf S}$ cannot be integrated over a finite volume of appropriate
dimension directly. The integration can be performed if before that one
contracts $\widetilde{\bf S}$ with a field of appropriate type, so that
their contraction would be a scalar-valued form. A more general possibility
is to construct the wedge product of $\widetilde{\bf S}$ with some other
form $\widetilde{\bf T}$ whose values are tensors complementary to those
which are the values of $\widetilde{\bf S}$ (i.e.\ tensors that can be
contracted with the value of $\widetilde{\bf S}$ to produce a scalar) and
then contract the values of $\widetilde{\bf S}$ and $\widetilde{\bf T}$.
In the follwoing this sort of expressions will be denoted as
\begin{displaymath}
\prec \widetilde{\bf S} \wedge \widetilde{\bf T} \succ.
\end{displaymath}
By definition, the contraction over the values does not affect those indices
of $\widetilde{\bf S}$ and $\widetilde{\bf T}$ over which they are contracted
with multivectors that characterize the infinitesimal integration volumes
and with respect to which they are antisymmetrized when one constructs their
wedge product. Thus, in the case of nonscalar-valued forms we will be dealing
with integrals of the form
\begin{displaymath}
\int_{V} \prec \widetilde{\bf S} \wedge \widetilde{\bf T} \succ.
\end{displaymath}

The calculus of nonscalar-valued five-vector forms is very similar to the
calculus of their four-vector counterparts. In view of this, it will be
convenient to recall first the basic definitions and formulae for forms
of the latter type and then to comment on the corresponding formulae
for five-vector forms.

An arbitrary nonscalar-valued four-vector $m$-form $\widetilde{\bf S}$ can
be presented as
\begin{displaymath}
\widetilde{\bf S} = \vec{\bf S}_{|\alpha_{1} \ldots \alpha_{m}|} \;
{\bf d}x^{\alpha_{1}} \wedge \ldots \wedge {\bf d}x^{\alpha_{m}},
\end{displaymath}
where the components $\vec{\bf S}_{\alpha_{1} \ldots \alpha_{m}}$ are
elements of some vector or tensor space. The wedge product of this $m$-form
with some other four-vector $n$-form
\begin{displaymath}
\widetilde{\bf T} = \vec{\bf T}_{|\alpha_{1} \ldots \alpha_{n}|} \;
{\bf d}x^{\alpha_{1}} \wedge \ldots \wedge {\bf d} x^{\alpha_{n}},
\end{displaymath}
whose values can be quantities of a totally different nature than those of
$\widetilde{\bf S}$, is the $(m+n)$-form
\begin{displaymath} \begin{array}{l}
\widetilde{\bf S} \wedge \widetilde{\bf T} = (\vec{\bf S}_{|\alpha_{1}
\ldots \alpha_{m}|} \otimes \vec{\bf T}_{|\alpha_{m+1} \ldots
\alpha_{m+n}|}) \\ \hspace{17ex} \rule{0ex}{3.5ex} \times \,
{\bf d} x^{\alpha_{1}} \wedge \ldots \wedge {\bf d} x^{\alpha_{m+n}}.
\end{array} \end{displaymath}
If the values of $\widetilde{\bf T}$ are tensors complementary to those which
are the values of $\widetilde{\bf S}$, one can contract $\widetilde{\bf S}$
and $\widetilde{\bf T}$ over the values and obtain the scalar-valued form
\begin{equation} \begin{array}{l}
\prec \widetilde{\bf S} \wedge \widetilde{\bf T} \succ \; = \; \prec
\vec{\bf S}_{|\alpha_{1} \ldots \alpha_{m}|} \, , \vec{\bf T}_{| \alpha_{m+1}
\ldots \alpha_{m+n}|} \succ \\ \hspace{19ex} \rule{0ex}{4ex} \times \,
{\bf d} x^{\alpha_{1}} \wedge \ldots \wedge {\bf d} x^{\alpha_{m+n}}.
\end{array} \end{equation}

According to definition (12) of part IV, the exterior derivative of form
(42) equals
\begin{displaymath} \begin{array}{l}
{\bf d} \! \prec \widetilde{\bf S} \wedge \widetilde{\bf T} \succ \, =
\partial_{\alpha} \! \prec \vec{\bf S}_{|\alpha_{1} \ldots \alpha_{m}|} \,
, \vec{\bf T}_{|\alpha_{m+1} \ldots \alpha_{m+n}|} \succ \\ \hspace{19ex}
\rule{0ex}{4ex} \times \, {\bf d} x^{\alpha} \wedge {\bf d}x^{\alpha_{1}}
\wedge \ldots \wedge {\bf d}x^{\alpha_{m+n}}.
\end{array} \end{displaymath}
By using the Leibniz rule for the operator $\nabla$, one can present this
derivative as
\begin{equation}
{\bf d} \! \prec \widetilde{\bf S} \wedge \widetilde{\bf T} \succ \, = \,
\prec {\bf d} \widetilde{\bf S} \wedge \widetilde{\bf T} \succ + \, (-1)^{m}
\! \prec \widetilde{\bf S} \wedge {\bf d} \widetilde{\bf T} \succ ,
\end{equation}
where, by definition,
\begin{equation}
{\bf d} \widetilde{\bf S} \equiv \nabla_{\alpha} \vec{\bf S}_{|\alpha_{1}
\ldots \alpha_{m}|} \; {\bf d} x^{\alpha} \wedge {\bf d} x^{\alpha_{1}}
\wedge \ldots \wedge {\bf d} x^{\alpha_{m}},
\end{equation}
and the derivative $\bf d \widetilde{T}$ is defined in a similar way. Form
(44) of rank $m+1$ is called the exterior derivative of the nonscalar-valued
four-vector $m$-form $\widetilde{\bf S}$.

  From the latter definition it follows that for an arbitrary nonscalar field
$\vec{\bf F}$ regarded as a four-vector 0-form, $\bf d \vec{F}$ is such a
nonscalar-valued 1-form that
\begin{displaymath} \bf
< d \vec{F} , U > \; = \nabla_{U} \vec{F}
\end{displaymath}
for any four-vector $\bf U$, which is the analog of formula (10) of
part IV. Using this relation, one can write:
\begin{displaymath}
{\bf d\widetilde{S}} = {\bf d} \vec{\bf S}_{|\alpha_{1} \ldots \alpha_{m}|}
\wedge {\bf d}x^{\alpha_{1}} \wedge \ldots \wedge {\bf d}x^{\alpha_{m}},
\end{displaymath}
which is the analog of formula (12) of part IV. The exterior derivative
of $\widetilde{\bf S}$ can also be defined in a manner similar to formulae
(14) of part IV: if $\widetilde{\bf S}$ is a 1-form, then
$\bf d \widetilde{S}$ is such a 2-form that
\begin{flushright} \hfill $\left. \begin{array}{l}
\bf <d\widetilde{S},U \wedge V> \; = \; \nabla_{U} <\widetilde{S},V>  \\
\bf \hspace*{10ex} - \; \nabla_{V}<\widetilde{S},U> - <\widetilde{S},[U,V]>
\end{array} \right.$ \hfill (45a) \end{flushright}
for any two four-vector fields $\bf U$ and $\bf V$; if $\widetilde{\bf S}$
is a 2-form, then $\bf d\widetilde{S}$ is such a 3-form that
\begin{flushright} $\left. \begin{array}{l}
\bf <d\widetilde{S},U \wedge V \wedge W> \\ \bf = \nabla_{U} <\widetilde
{S}, V \wedge W> + \; \nabla_{V} <\widetilde{S},W \wedge U> \\ \bf + \;
\nabla_{W} <\widetilde{S},U \wedge V> - <\widetilde{S}, [U,V] \wedge  W>
\\ \bf - <\widetilde{S},[V,W] \wedge U> - <\widetilde{S},[W,U] \wedge V>
\end{array} \right.$ (45b) \end{flushright} \setcounter{equation}{45}
for any four-vector fields $\bf U$, $\bf V$ and $\bf W$; etc. It is not
difficult to see that for nonscalar-valued forms one can use formula (13a)
of part IV for the exterior derivative of a wedge product. However, in the
general case the double derivative $\bf dd \widetilde{S}$ of an arbitrary
nonscalar-valued form $\widetilde{\bf S}$ is not identically zero.

Everything that has be said above about four-vector forms applies, with
obvious modification, to five-vector forms as well. An essentially new
feature here is that for an arbitrary nonscalar-valued five-vector $m$-form
\begin{equation}
\widetilde{\bf t} = \vec{\bf t}_{|A_{1} \ldots A_{m}|} \; \widetilde
{\bf o}^{A_{1}} \wedge \ldots \wedge \widetilde{\bf o}^{A_{m}}
\end{equation}
(${\bf o}^{A}$ is a basis of five-vector 1-forms dual to some passive regular
coordinate five-vector basis) one can define {\em two} exterior derivatives:
\begin{equation}
\nabla_{A} \vec{\bf t}_{|A_{1} \ldots A_{m}|} \; \widetilde{\bf o}^{A} \wedge
\widetilde{\bf o}^{A_{1}} \wedge \ldots \wedge \widetilde{\bf o}^{A_{m}}
\end{equation}
and
\begin{equation}
\plaision_{A}\vec{\bf t}_{|A_{1}\ldots A_{m}|} \; \widetilde{\bf o}^{A}\wedge
\widetilde{\bf o}^{A_{1}} \wedge \ldots \wedge \widetilde{\bf o}^{A_{m}}.
\end{equation}
Both of these derivatives are well-defined and both may be of use in
calculations. In the following I will use the symbol $\bf d$ to denote
derivative (48) and will refer to it simply as to the exterior derivative
of $\widetilde{\bf t}$. For derivative (47) I will introduce the notation
${\bf d}^{\scriptscriptstyle \nabla}$. It is easy to see that for any
$\widetilde{\bf t}$ the derivatives ${\bf d} \widetilde{\bf t}$ and
${\bf d}^{\scriptscriptstyle \nabla} \widetilde{\bf t}$ differ by
$({\bf d} \widetilde{\bf t}^{\cal Z})^{\cal E}$.

It is not difficult to obtain for $\bf d$ and ${\bf d}^{\scriptscriptstyle
\nabla}$ the analogs of formulae (19) and (21) of part IV, where instead
of operators $\partial$ there will now stand operators $\plaision$ and
$\nabla$, respectively. Formulae (20) and (22a) of part IV will apply
to nonscalar-valued forms without any changes, however, in the general case,
identity (22b) will no longer be valid.

\vspace{3ex} \begin{flushleft}
E. \it Five-vector analogs of the field strength tensor
\end{flushleft}
As is known, the action of the operator $\bf dd$ on an arbitrary
nonscalar-valued four-vector form $\widetilde{\bf S}$ of rank less than 3
does not produce an identical zero. Instead, it produces a form which is
the wedge product of $\widetilde{\bf S}$ with a certain 2-form, $\RR$,
whose values are linear algebraic operators that act upon the values of
$\widetilde{\bf S}$. The operator-valued 2-form $\RR$ is the same for all
forms whose values belong to the same vector or tensor space, and does not
depend on the rank of the form. If the values of $\widetilde{\bf S}$ are
elements of the tensor product of vector spaces $W_{1},W_{2},\ldots,W_{n}$,
then the 2-form $\RR$ corresponding to $\widetilde{\bf S}$ can be constructed
according to a simple recipe from the 2-forms $\RR$ that correspond to the
forms whose values are elements of $W_{1}, W_{2}, \ldots, W_{n}$. If the
values of $\widetilde{\bf S}$ are linear forms associated with some vector
space $W$, then the corresponding 2-form $\RR$ differs only in the sign and
transposition from the 2-form $\RR$ that corresponds to those forms whose
values are elements of $W$. Finally, for the scalar-valued forms $\RR$ is
identically zero, which manifests itself in identity (13b) of part IV.

We thus see that to be able to evaluate the double exterior derivatives of
arbitrary nonscalar-valued forms it is sufficient to know the operator-valued
2-forms $\RR$ only for the 0-forms and only for those of them whose values
are {\em vectors} (not tensors). As is known, the 2-form $\RR$ that
corresponds to four-vectors is called the curvature tensor and is denoted
as $\bf R$. The 2-forms $\RR$ that correspond to nonspacetime vectors are
called field strength tensors and are usually denoted as $\bf F$ or $\bf G$.

Everything that has been said above is true of nonscalar-valued five-vector
forms as well. In this section I will consider the five-vector analogs of
the field strength tensors and will discuss their basic properties. The
five-vector analog of the curvature tensor will be discussed in the next
part of the series.

Since in the case of nonscalar-valued five-vector forms one has {\em two}
exterior derivative operators, ${\bf d}^{\scriptscriptstyle \nabla}$ and
$\bf d$, one can consider two five-vector analogs of the field strength
tensor, which will be denoted as ${\bf F}^{\scriptscriptstyle \nabla}$ and
$\bf F$, respectively. As in section C, let us consider a certain set of
fields whose values are some nonspacetime vectors. Regarding these fields
as nonscalar-valued five-vector 0-forms, for an arbitrary field
$\smallvec{S}$ one has
\begin{displaymath}
< {\bf d}^{\scriptscriptstyle \nabla} \smallvec{S} \, , {\bf w} > \;
= \nabla_{\bf w} \smallvec{S}.
\end{displaymath}
Substituting this expression into the analog of formula (45a) for
five-vector forms, one finds that for any two five-vector fields
$\bf u$ and $\bf v$
\begin{displaymath} \begin{array}{l}
< {\bf d}^{\scriptscriptstyle \nabla} {\bf d}^{\scriptscriptstyle \nabla}
\smallvec{S} \, , {\bf u \wedge v} > \; = \nabla_{\bf u}
< {\bf d}^{\scriptscriptstyle \nabla} \smallvec{S} \, , {\bf v} > \\
\rule{0ex}{3ex}\hspace{13ex} - \, \nabla_{\bf v}
<{\bf d}^{\scriptscriptstyle \nabla} \smallvec{S} \, ,{\bf u}> -
<{\bf d}^{\scriptscriptstyle \nabla} \smallvec{S} \, , [{\bf u,v}]> \\
\rule{0ex}{3ex}\hspace{18.7ex} = (\nabla_{\bf u} \nabla_{\bf v} -
\nabla_{\bf v} \nabla_{\bf u} - \nabla_{\bf [u,v]}) \, \smallvec{S} \\
\rule{0ex}{3ex}\hspace{18.7ex} \equiv \; {\bf F}^{\scriptscriptstyle \nabla}
({\bf u \wedge v}) \, \smallvec{S}.
\end{array} \end{displaymath}
It is evident that operator ${\bf F}^{\scriptscriptstyle \nabla}({\bf u
\wedge v})$ depends only on the $\cal Z$-components of $\bf u$ and $\bf v$.
If $\bf u \in U$ and $\bf v \in V$, then $\nabla_{\bf u} = \nabla_{\bf U}$,
$\nabla_{\bf v} = \nabla_{\bf V}$ and $\nabla_{\bf [u,v]} =
\nabla_{\bf [U,V]}$, and the right-hand side will equal
\begin{displaymath}
(\nabla_{\bf U} \nabla_{\bf V} - \nabla_{\bf V} \nabla_{\bf U}
 - \nabla_{\bf [U,V]}) \, \smallvec{S},
\end{displaymath}
which is the usual expression for the four-vector field strength tensor, so
${\bf F}^{\scriptscriptstyle \nabla} ({\bf u \wedge v})$ is simply the
five-vector equivalent of the latter. Among other things this means that
${\bf F}^{\scriptscriptstyle \nabla}$ is a local operator. The result of
its action on an arbitrary field $\smallvec{S} = s^{i} \smallvec{E}_{i}$
is expressed in components as
\begin{equation}
( \, {\bf F}^{\scriptscriptstyle \nabla} \smallvec{S} \, )^{i}_{\; AB} =
F^{({\scriptscriptstyle \nabla}) \, i}_{\hspace{3ex} jAB} \, s^{j},
\end{equation}
and it is not difficult to see that
\begin{equation}
F^{({\scriptscriptstyle \nabla}) \, i}_{\hspace{3ex} j \alpha 5} =
F^{({\scriptscriptstyle \nabla}) \, i}_{\hspace{3ex} j5 \alpha} = 0
\end{equation}
and that in any standard coordinate basis
\begin{equation} \begin{array}{rcl}
F^{({\scriptscriptstyle \nabla}) \, i}_{\hspace{3ex} j \alpha \beta}
& = & \partial_{\alpha} A^{i}_{\; j \beta} - \partial_{\beta}
A^{i}_{\; j \alpha} \\ & & + \; A^{i}_{\; k \alpha} A^{k}_{\; j \beta}
- A^{i}_{\; k \beta} A^{k}_{\; j \alpha}, \rule{0ex}{2.5ex}
\end{array} \end{equation}
where $A^{i}_{\; j \alpha}$ are the gauge fields that correspond to the
basis fields $\smallvec{E}_{i}$.

Performing the same calculations with the operator $\bf d$, one obtains that
\begin{displaymath} \begin{array}{l}
< {\bf d d} \smallvec{S} \, , {\bf u \wedge v} > \\ \hspace{3ex} =
(\plaision_{\bf u} \plaision_{\bf v} - \plaision_{\bf v} \plaision_{\bf u}
- \plaision_{\bf [u,v]}) \, \smallvec{S} \; \equiv \;
{\bf F} ({\bf u \wedge v}) \, \smallvec{S} \rule{0ex}{3ex} ,
\end{array} \end{displaymath}
where $\bf F (u \wedge v)$ can be shown to be a linear algebraic operator
and is a contraction of a certain operator-valued 2-form $\bf F$ with the
bivector $\bf u \wedge v$. An equivalent expression for $\bf F$ can be
obtained by using definition (48). In components, the effect of $\bf F$
on an arbitrary 0-form $\smallvec{S}$ is given by a formula similar to
formula (49):
\begin{displaymath}
( \, {\bf F} \smallvec{S} \, )^{i}_{\; AB} = F^{\, i}_{\; jAB} \, s^{j},
\end{displaymath}
and instead of (50) and (51) one now has
\begin{equation} \begin{array}{rcl}
F^{\, i}_{\; jAB} & = & \partial_{A} B^{i}_{\; jB}
- \partial_{B} B^{i}_{\; jA} \\ & & + \; B^{i}_{\; kA} B^{k}_{\; jB}
- B^{i}_{\; kB} B^{k}_{\; jA}, \rule{0ex}{2.5ex}
\end{array} \end{equation}
where $B^{i}_{\; jA}$ are the five-vector gauge fields that correspond to
the basis fields $\smallvec{E}_{i}$. From the latter formula it follows
that in any regular five-vector basis
\begin{displaymath}
F^{\, i}_{\; j \alpha \beta} =
F^{({\scriptscriptstyle \nabla}) \, i}_{\hspace{3ex} j \alpha \beta},
\end{displaymath}
which is a reflection of the fact that the 2-form
${\bf F}^{\scriptscriptstyle \nabla}$ equals the
$\widetilde{\cal Z}$-component of $\bf F$.

Each five-vector field strength tensor has an important differential
property, which directly follows from its definition:
\begin{equation}
\bf dF = 0,
\end{equation}
where $\bf F$ is regarded as a 2-form whose values are nonspacetime tensors
of rank $(1,1)$. There are several equivalent ways of proving equation (53).
For example, one can present the field produced by the action of $\bf F$ on
an arbitrary 0-form $\smallvec{S}$ as $\prec {\bf F} \wedge \smallvec{S}
\succ$, where it is assumed that the value of $\smallvec{S}$ is contracted
with the value of $\bf F$ as with a tensor, and then consider the quantity
$\bf ddd \smallvec{S}$ and evaluate it in two different ways. On the one
hand, $\bf ddd \smallvec{S} = \bf dd (d \smallvec{S})$, and since the action
of $\bf dd$ is independent of the rank of the form, and the values of
${\bf d} \smallvec{S}$ and $\smallvec{S}$ are elements of the same vector
space, one has
\begin{equation}
\bf ddd \smallvec{S} = \; \prec F \wedge d \smallvec{S} \succ.
\end{equation}
On the other hand, $\bf ddd \smallvec{S} = d(dd \smallvec{S}) = d \! \prec F
\wedge \smallvec{S} \succ$, and by using formula (43) one obtains
\begin{equation}
\bf ddd \smallvec{S} = \;
\prec d F \wedge \smallvec{S} \succ +  \prec F \wedge d \smallvec{S} \succ.
\end{equation}
Comparing expressions (54) and (55) and considering that $\smallvec{S}$ is
an arbitrary 0-form, one concludes that the derivative $\bf dF$ should be
identically zero.

Equation (53) can also be proved by using definition (48) and the
Jacobi identity in application to the commutators of operators
$\plaision$, or by straightforward calculation with components.
Since for any form $\widetilde{\bf t}$ one has $({\bf d}
\widetilde{\bf t})^{\widetilde{\cal Z}} = {\bf d}^{\scriptscriptstyle \nabla}
\widetilde{\bf t}^{\widetilde{\cal Z}}$, from equation (53) follows a similar
differential identity for ${\bf F}^{\scriptscriptstyle \nabla}$:
\begin{displaymath}
\bf d^{\scriptscriptstyle \nabla} F^{\scriptscriptstyle \nabla} = 0,
\end{displaymath}
which, naturally, can be proved in an independent way.

\vspace{3ex} \begin{flushleft}
F. \it Nonspacetime analogs of five-vectors
\end{flushleft}
The nonspacetime vectors which we have talked about so far and which are
used in physics, for example, for describing the internal symmetries of
elementary particles, resemble ordinary tangent vectors in the sense that at
each space-time point their vector space is endowed only with a nondegenerate
inner product and has no other additional structure similar to the
$\cal Z$--$\cal E$ splitting in the space of five-vectors. In accordance
with this, on the parallel transport of such vectors one imposes no other
constraints except for the requirements that it be linear and conserve the
mentioned inner product, so at an appropriate choice of the relevant gauge
fields, any given vector at the initial point can be transported into any
vector of the same length at the final point, if this does not contradict
the condition of the transport continuity. One may now ask the following
question: can there be defined such nonspacetime vectors that would resemble
five-vectors?

Let us try to imagine what properties such vectors should have. It goes
without saying that at each space-time point they should form a certain
finite-dimensional vector space, $W$, the dimension of which in the general
case it is convenient to denote as $n+1$. Accordingly, in the following such
vectors themselves will be referred to as $(n+1)$-{\em vectors}, and will be
denoted with lower-case Roman-type letters with an arrow: $\vec{\rm u}$,
$\vec{\rm v}$, $\vec{\rm w}$, etc. It is also natural to assume that the
space of $(n+1)$-vectors is endowed with some nondegenerate inner product,
which will be denoted as $\eta$. All this, however, applies to ordinary
nonspacetime vectors as well. It seems reasonable to suppose that
$(n+1)$-vectors differ from the latter in that their space is ``split''
into two invariant subspaces, which will be denoted as $W^{\cal Z}$ and
$W^{\cal E}$, the first one of dimension $n$, the other one-dimensional.
The space $W$ itself will be the direct sum of these two subspaces, and
the components of an arbitrary $(n+1)$-vector $\vec{\rm u}$ in them will be
denoted as $\vec{\rm u}^{\cal Z}$ and $\vec{\rm u}^{\cal E}$, respectively.

Since as in the case of ordinary nonspacetime vectors, it is not supposed
that $(n+1)$-vectors are associated with any manifold, the mentioned
splitting will have a real meaning only if it manifests itself in some
specific properties, basing on which one would be able to say that one is
dealing with $(n+1)$-vectors and not with some type of ordinary nonspacetime
vectors of dimension $n+1$. It is apparent that if the space of
$(n+1)$-vectors is not endowed with any additional structure, then the
above specific properties can only be related to parallel transport. Basing
on the analogy with five-vectors, let us assume that $(n+1)$-vectors from
$W^{\cal E}$ are transported into $(n+1)$-vectors from $W^{\cal E}$
and that $(n+1)$-vectors from $W^{\cal Z}$ may acquire in the process of
transport a nonzero $\cal E$-component. The first of these properties tells
us that we are not dealing with ordinary nonspacetime vectors. The second
property tells us that neither we are dealing with elements of the direct
sum of two spaces of ordinary nonspacetime vectors (of dimension $n$ and
one). In addition to this, let us assume that parallel transport conserves
the inner product
\begin{displaymath}
\theta ( \vec{\rm u}, \vec{\rm v} ) \; \equiv \;
\eta ( \vec{\rm u}^{\cal Z}, \vec{\rm v}^{\cal Z} ),
\end{displaymath}
which is the analog of the inner product $g$ for five-vectors.

In order to write down the indicated properties of $(n+1)$-vectors in the
form of equations, let us introduce the following notations. The set of all
sufficiently smooth fields whose values are $(n+1)$-vectors of the considered
type will be denoted as $\WW$. An arbitrary set of basis fields from $\WW$
will be denoted as $\vec{\rm e}_{1}$, \ldots , $\vec{\rm e}_{n+1}$. In those
cases where it cannot lead to confusion with similar notations for the
analogs of five-vectors in three-dimensional Euclidean space, it will be
taken that lower-case latin indices run 1 through $n$, while capital Greek
indices will be assumed to run 1 through $n+1$. Often, instead of the value
$n+1$ I will use the symbol $\&$.

The basis in $\WW$ can be chosen arbitrarily. However, for practical reasons
it is more convenient to select it in such a way that at each space-time
point the $(n+1)${\em st} basis vector belong to $W^{\cal E}$. Similarly to
the case of five-vectors, such bases will be called {\em standard}. It is
also useful to introduce the notion of a {\em regular} basis, whose first $n$
elements belong to $W^{\cal Z}$ and the $(n+1)${\em st} element is normalized
in some particular way. Since $(n+1)$-vectors are not associated with any
manifold, and therefore cannot be represented with differential-algebraic
operators, and since, as one will see below, from the rules of their parallel
transport one also cannot obtain any special normalization for the
vectors from $W^{\cal E}$, the only condition that one can use for
normalizing $\vec{\rm e}_{\&}$ is the requirement $| \eta (\vec{\rm e}_{\&},
\vec{\rm e}_{\&}) | = 1$, which is similar to the normalization condition
for the fifth basis vector in a normalized regular five-vector basis.

The connection coefficients for an arbitrary set of basis fields
$\vec{\rm e}_{\Theta}$ in $\WW$ are defined in the usual way:
\begin{displaymath}
\plaision_{A} \vec{\rm e}_{\Theta} =
\vec{\rm e}_{\Xi} \, C^{\Xi}_{\; \; \Theta A}.
\end{displaymath}
The quantities $C^{\Xi}_{\; \; \Theta A}$ will still be called
{\em five-vector gauge fields}. From the assumptions made above about the
parallel transport of $(n+1)$-vectors it follows that for any standard basis
\begin{equation}
C^{i}_{\; \& A} = 0,
\end{equation}
which is the analog of constraint (20) on the connection coefficients
for five-vector fields. Furthermore, if, for example, the considered
$(n+1)$-vectors are complex and their inner product $\theta$ is Hermitian,
there should hold the equation
\begin{equation}
\partial_{A} \theta_{ij} - \theta_{kj} (C^{\, k}_{\, \; iA})^{\ast}
- \theta_{ik} C^{\, k}_{\, \; jA} = 0,
\end{equation}
similar to constraint (41) on the five-vector gauge fields associated with
ordinary nonspacetime vectors.

  From the assumptions made above it follows that parallel transport of
$(n+1)$-vectors preserves the following equivalence relation on $W$:
\begin{displaymath}
\vec{\rm u} \equiv \vec{\rm v} \; \Leftrightarrow \;
\vec{\rm u} - \vec{\rm v} \in W^{\cal E}.
\end{displaymath}
It is a simple matter to check that with regard to their properties, the
elements of the quotient space $W/W^{\cal E}$ are ordinary nonspacetime
vectors, and that at each space-time point this quotient space, endowed with
the inner product induced by the product $\theta$ on $W$, is isomorphic to
the subspace $W^{\cal Z}$. One should therefore expect that with each type of
$(n+1)$-vectors there is associated a certain type of ordinary nonspacetime
vectors, whose relation to the considered $(n+1)$-vectors is similar to the
relation of four-vectors to five-vectors. For these associated vectors one
can use all the notations and definitions that have been introduced earlier
for ordinary nonspacetime vectors. In particular, if the gauge fields
corresponding to them are defined by equation (39) and if the corresponding
basis fields $\smallvec{E}_{i}$ are such that at each point
$\smallvec{E}_{i}$ is the equivalence class of the basis $(n+1)$-vector
$\vec{\rm e}_{i}$, then by virtue of what has been said above there should
hold the equation
\begin{equation}
C^{i}_{\; jA} = B^{i}_{\; jA},
\end{equation}
which is the analog of relation (48) of part II between the connection
coefficients for four-vector and five-vector fields.

Let us now consider more closely the gauge fields $C^{\Xi}_{\; \; \Theta A}$.
First of all, let us write down the formula for their transformation as one
passes to another set of basis fields in $\WW$:
\begin{equation}
C'^{\, \Theta}_{\; \; \, \Xi A} = (L^{-1})^{\Theta}_{\; \Delta} \,
C^{\Delta}_{\; \; \Sigma A} \, L^{\Sigma}_{\; \Xi} +
(L^{-1})^{\Theta}_{\; \Delta} \, \partial_{A} L^{\Delta}_{\; \Xi} ,
\end{equation}
where $L^{\Xi}_{\; \Theta}$ is the basis transformation matrix. If both
bases are standard, one has $L^{i}_{\; \&} = (L^{-1})^{i}_{\; \&} = 0$,
and at $\Theta = i$ and $\Xi = \&$ obtains
\begin{displaymath} \begin{array}{l}
C'^{\, i}_{\; \; \, \& A} = (L^{-1})^{i}_{\; k} \, C^{k}_{\; \; l A} \,
L^{l}_{\; \&} \\ \hspace*{5ex} + \; (L^{-1})^{i}_{\; k} \,
C^{k}_{\; \; \& A} \, L^{\&}_{\; \&} + (L^{-1})^{i}_{\; k} \,
\partial_{A} L^{k}_{\; \&} = 0, \rule{0ex}{3ex}
\end{array} \end{displaymath}
which is actually a demonstration of the fact that from the validity of
equation (56) in one standard basis follows its validity in any other
such basis. In a similar manner, at $\Theta = i$ and $\Xi = j$ one has
\begin{displaymath}
C'^{\, i}_{\; \; \, j A} =
(L^{-1})^{i}_{\; k} \, C^{k}_{\; \; l A} \, L^{l}_{\; j} +
(L^{-1})^{i}_{\; k} \, \partial_{A} L^{k}_{\; j},
\end{displaymath}
so the connection coefficients $C^{\, i}_{\; \; j A}$ transform as gauge
fields corresponding to ordinary nonspacetime vectors, which agrees with
equation (58).

Let us now turn to the gauge fields that determine the $\cal E$-components
of the transported $(n+1)$-vectors. The first question one has to ask
is whether parallel transport conserves the length of the vectors from
$W^{\cal E}$. Since $(n+1)$-vectors are not associated with any manifold,
the only measure available for the vectors from $W^{\cal E}$ is the scalar
square constructed with the inner product $\eta$. As in the case of
five-vectors, let us suppose that this scalar square does {\em not} change.
In the case of real vectors this means that for any field of regular bases
one should have $C^{\&}_{\; \& A} = 0$. In the case of complex vectors and
Hermitian $\eta$, the fields $C^{\&}_{\; \& A}$ for a regular basis do not
have to vanish, and it is only necessary that they be imaginary. With
transition to another basis in $\WW$, but also a regular one, in the
latter case one has $L^{\&}_{\; \&} = e^{i \alpha}$, so
\begin{displaymath} \begin{array}{rcl}
C'^{\, \&}_{\; \; \; \& A} & = & (L^{-1})^{\&}_{\; \&} \, C^{\&}_{\; \; \& A}
\, L^{\&}_{\; \&} + (L^{-1})^{\&}_{\; \&} \, \partial_{A} L^{\&}_{\; \&} \\
& = & C^{\&}_{\; \; \& A} + i \, \partial_{A} \alpha . \rule{0ex}{3ex}
\end{array} \end{displaymath}

There is one more constraint that can be imposed on the parallel transport
of $(n+1)$-vectors, which implicitly is very often imposed on the parallel
transport of ordinary nonspacetime vectors. Namely, one can require that
this transport conserve the Levi-Civita type tensor $\epsilon$ associated
with the considered $(n+1)$-vectors. In the case of real $W$ this condition
is equivalent to the conservation of the length of the $(n+1)$-vectors from
$W^{\cal E}$. In the case of complex $W$ this requirement can be shown to
imply that in any basis where the components of $\eta$ and $\epsilon$ are
constant, one should have $C^{\Theta}_{\; \Theta A} = 0$. What the latter
equation results in will be seen below.

The gauge fields $C^{\&}_{\; \; j A}$ are evidently the analogs of the
five-vector connection coefficients $H^{5}_{\, \mu A}$. From formula (59)
it follows that with transition to another basis in $\WW$ they transform as
\begin{displaymath} \begin{array}{l}
C'^{\, \&}_{\; \; \; j A} =
(L^{-1})^{\&}_{\; \Xi} \, C^{\Xi}_{\; \; \, l A} \, L^{l}_{\; j} +
(L^{-1})^{\&}_{\; \&} \, C^{\&}_{\; \; \& A} \, L^{\&}_{\; j} \\
\hspace*{9ex} + \; (L^{-1})^{\&}_{\; k} \, \partial_{A} L^{k}_{\; j} +
(L^{-1})^{\&}_{\; \&} \, \partial_{A} L^{\&}_{\; j}. \rule{0ex}{3ex}
\end{array} \end{displaymath}
If both bases are regular, then $(L^{-1})^{\&}_{\; j} = L^{\&}_{\; j} = 0$,
and one has
\begin{equation}
C'^{\, \&}_{\; \; \; j A} = (L^{-1})^{\&}_{\; \&} \,
C^{\&}_{\; \; l A} \, L^{l}_{\; j}.
\end{equation}
If, in addition, one has $\vec{\rm e}^{ \, \prime}_{\&} =
\vec{\rm e}_{\&}$, then simply
\begin{displaymath}
C'^{\, \&}_{\; \; \; j A} = C^{\&}_{\; \; l A} \, L^{l}_{\; j}.
\end{displaymath}

An essential difference between the gauge fields $C^{\&}_{\; \; j A}$ and
their five-vector counterparts is that for the former there does not exist
a nonzero value that would be invariant under the transformations from the
symmetry group of $W$. On the other hand, the value $C^{\&}_{\; \; j A} = 0$,
which does not break this symmetry, has the unpleasant property that at it
one cannot distinguish the considered $(n+1)$-vectors from pairs made of an
ordinary $n$-dimensional nonspacetime vector and a scalar. It is evident that
at any nonzero $C^{\&}_{\; \; j A}$ the inner product $\eta$ is not conserved
by parallel transport, and since neither the requirement of the covariant
constancy of $\theta$ nor a similar requirement for the $n$-plus-one-vector
$\epsilon$ tensor impose any constraints on $C^{\&}_{\; \; j A}$, the latter
can be absolutely arbitrary.

Let us examine in more detail the case of complex vectors for which the
inner product $\eta$ is Hermitian and is positively definite. At each
space-time point, let us select the basis in $W$ orthonormal and such
that one would have $\epsilon_{1 \ldots n {\scriptscriptstyle \&}} = 1$.
Condition (57) will then acquire the form
\begin{displaymath}
\theta_{kj}(C^{\, k}_{\, \; iA})^{\ast} + \theta_{ik}C^{\, k}_{\, \; jA} = 0,
\end{displaymath}
whence it follows that the quantities $C_{ijA} \equiv \theta_{ik}
C^{\, k}_{\, \; jA}$ are anti-Hermitian matrices with respect to the indices
$i$ and $j$. Since in the selected basis $C_{ijA} = C^{\, i}_{\, \; jA}$,
one can write that
\begin{equation} \begin{array}{l}
C^{\, i}_{\, \; jA} = (i/2) \, g \, (t_{a})^{i}_{\, j} \, C^{a}_{\; A} \\
\hspace*{10ex} + \; ig \, [2n(n+1)]^{-1/2} \, \delta^{i}_{\, j} \,
C^{\scriptscriptstyle 0}_{\, A}, \rule{0ex}{3ex}
\end{array} \end{equation}
where the index $a$ runs 1 through $n^{2} \! - \! 1$; the matrices
$(t_{a})^{i}_{\, j}$ are the usual (Hermitian) generators for the
fundamental representation of SU($n$), normalized by the condition
${\rm Tr}(t_{a} t_{b}) = 2 \delta_{ab}$; the fields $C^{a}_{\; A}$ and
$C^{\scriptscriptstyle 0}_{\, A}$ are real; and $g$ is a dimensionless
constant, which together with the factors $1/2$ and $[2n(n+1)]^{-1/2}$ is
introduced for convenience. From the condition $C^{\Theta}_{\; \Theta A} =
0$ it follows that
\begin{equation}
C^{\, \&}_{\; \; \& A} = - \, ig \, [n/2(n+1)]^{1/2} \,
C^{\scriptscriptstyle 0}_{\, A}.
\end{equation}
By using (61) and (62) one can write down the expression for the
components of the five-vector covariant derivative of an arbitrary
$(n+1)$-vector field in the selected basis in the following way:
\begin{flushright}
\hspace*{0ex} \hfill $ \left. \begin{array}{l}
(\plaision_{A} \vec{\rm u})^{i} = \partial_{A} u^{i} + (i/2) \, g \,
(t_{a})^{i}_{\, j} \, C^{a}_{\; A} \, u^{j} \\ \hspace*{17ex} + \; ig \,
[2n(n+1)]^{-1/2} \, C^{\scriptscriptstyle 0}_{\, A} \, u^{i}
\end{array} \right. $ \hfill (63a) \\
\hspace*{0ex} \hfill $ \left. \begin{array}{l}
(\plaision_{A} \vec{\rm u})^{\&} = \partial_{A} u^{\&}
\rule{0ex}{4ex} \\ \hspace*{4ex} - \; ig \, [n/2(n+1)]^{1/2} \,
C^{\scriptscriptstyle 0}_{\; A} \, u^{\&} + g X_{jA} \, u^{j},
\end{array} \right. $ \hfill (63b)
\end{flushright} \setcounter{equation}{63}
where there has been introduced the notation $X_{jA} \equiv g^{-1}
C^{\&}_{\; \; j A}$. Similarly, the expression for the components of the
five-vector covariant derivative of a field $\widetilde{\rm v}$ whose
values are elements of the space $\widetilde{W}$ of linear forms on $W$
can be written down as follows:
\begin{flushright}
$ \left. \begin{array}{l}
(\plaision_{A} \widetilde{\rm v})_{i} = \partial_{A} v_{i} - (i/2) \, g \,
v_{j} \, (t_{a})^{j}_{\, i} \, C^{a}_{\; A} \\ \hspace*{7ex} - \; ig \,
[2n(n+1)]^{-1/2} \, v_{i} \, C^{\scriptscriptstyle 0}_{\, A} - g v_{\&}
X_{iA} \end{array} \right. $ \hfill (64a) \\ $ \left. \begin{array}{l}
(\plaision_{A} \widetilde{\rm v})_{\&} = \partial_{A} v_{\&} +
ig \, [n/2(n+1)]^{1/2} \, v_{\&} C^{\scriptscriptstyle 0}_{\, A}.
\end{array} \right. $ \hfill \rule{0ex}{4ex}(64b)
\end{flushright} \setcounter{equation}{64}

If one disregards the terms involving the fields $X_{iA}$, the expressions
obtained will have such a form as if one was dealing with the gauge fields
corresponding to ordinary nonspacetime vectors and the gauge group was
${\rm SU}(n) \times {\rm U}(1)$. With respect to ${\rm SU}(n)$ the sets of
fields $(u^{1}, \ldots ,u^{n})$ and $(v_{1}, \ldots , v_{n})$ transform
according to the fundamental and anti-fundamental representations,
respectively, and the fields $u^{\&}$ and $v_{\&}$ are singlets.
With respect to the group ${\rm U}(1)$ the fields $u^{1}, \ldots , u^{n}$
all have the charge $g[2n(n+1)]^{-1/2}$, the field $u^{\&}$ has the
charge $-g[n/2(n+1)]^{1/2}$, and the charges of the fields $v_{1}, \ldots ,
v_{\&}$ are opposite to those of $u^{1},\ldots, u^{\&}$, which is in
agreement with the fact that the field $\widetilde{\rm v}$ can be
obtained from some $(n+1)$-vector field by conjugation (by the latter I mean
the antilinear map from $W$ to $\widetilde{W}$ fixed by the inner product
$\eta$, which is the analog of the map $\vartheta_{h}$ considered in section
3 of part II, and which in the selected basis coincides with ordinary
Hermitian conjugation).

Besides $C^{a}_{\; A}$ and $C^{\scriptscriptstyle 0}_{\, A}$, the above
expressions for the derivatives involve the gauge fields $X_{jA}$, due to
which the covariant differentiation of the considered $(n+1)$-vector fields
in general does not commute with conjugation, as it can be clearly seen by
comparing formulae (63) and (64). To gain a better understanding of what
this noncommutativity implies, let us recall how one assigns a representation
to matter fields in ordinary gauge theory when introducing new gauge fields.
To be definite, let us consider the case where the gauge group in question
is unitary. As always, the starting point is the existence of several matter
fields in the theory, say, $\varphi^{1}, \ldots , \varphi^{n}$, that enter
the Lagrangian density in such a way that the latter is invariant under the
replacement
\begin{equation}
\varphi^{i} \rightarrow \varphi^{\prime \, i} = L^{i}_{\; j} \, \varphi^{j},
\end{equation}
where $L^{i}_{\; j}$ is an arbitrary constant unitary $n \times n$ matrix.
One then gauges this symmetry by introducing the corresponding gauge fields,
and as a result obtains the following expression for the derivative of the
set $\varphi = (\varphi^{1}, \ldots , \varphi^{n})$:
\begin{displaymath}
(\nabla_{\alpha} \varphi )^{i} = \partial_{\alpha} \varphi^{i} +
(i/2) \, g \, (t_{a})^{i}_{\, j} \, B^{a}_{\; \alpha} \, \varphi^{j} +
ig' B^{\scriptscriptstyle 0}_{\; \alpha} \, \varphi^{i},
\end{displaymath}
where $(t_{a})^{i}_{\, j}$ are the same as in formula (61), and for
simplicity I omit the connection coefficients corresponding to other
possible degrees of freedom of $\varphi$. By presenting the transformation
formula for the considered matter fields in the form (65) one thereby
states that this set of fields transforms according to the {\em fundamental}
representation of the gauge group (= these fields are components of a
corresponding nonspacetime {\em vector}). Equally well, one can lable the
fields with a {\em lower} index and, accordingly, write the rule for their
transformation as
\begin{displaymath}
\varphi_{i} \rightarrow \varphi'_{i} = \varphi_{j} \, L^{j}_{\; i}.
\end{displaymath}
By doing so one would state that the fields $\varphi$ transform according
to the {\em anti-fundamental} representation (= are components of a
{\em linear form} associated with the relevant nonspacetime vectors), and
the expression for the derivative would then acquire the form
\begin{displaymath}
(\nabla_{\alpha} \varphi )_{i} = \partial_{\alpha}\varphi_{i} - (i/2)
\, g \, \varphi_{j} \, (t_{a})^{j}_{\, i} \, \widetilde{B}^{a}_{\; \alpha}
- ig' \varphi_{i} \, \widetilde{B}^{\scriptscriptstyle 0}_{\; \alpha},
\end{displaymath}
where $\varphi_{i} = \varphi^{i}$, $\widetilde{B}^{\scriptscriptstyle 0}_
{\; \alpha} = - B^{\scriptscriptstyle 0}_{\; \alpha}$, $\widetilde{B}^{a}_
{\; \alpha} = - \varepsilon^{a}_{b} B^{b}_{\; \alpha}$, and the coefficients
$\varepsilon^{a}_{b}$ are determined by the equation $(t_{a})^{i}_{\, j}
= (t_{b})^{j}_{\, i} \, \varepsilon^{b}_{a}$. The transition from the
fields $\varphi^{i},B^{\scriptscriptstyle 0}_{\; \alpha}, B^{a}_{\; \alpha}$
to the fields $\varphi_{i}, \widetilde{B}^{\scriptscriptstyle 0}_{\; \alpha},
\widetilde{B}^{a}_{\; \alpha}$ and vice versa is a part of the charge
conjugation operation.

By making similar transformations in formulae (63) and (64) one obtains
\begin{flushright} $ \left. \begin{array}{l}
(\plaision_{A} \vec{\rm u})_{i}=\partial_{A} u_{i} - (i/2) \, g \, u_{j} \,
(t_{a})^{j}_{\, i} \, \widetilde{C}^{a}_{\; A} \, \\ \hspace*{17ex} - \; ig
\, [2n(n+1)]^{-1/2} \, u_{i} \, \widetilde{C}^{\scriptscriptstyle 0}_{\, A}
\end{array} \right. $ \hfill (66a) \\ $ \left. \begin{array}{l}
(\plaision_{A} \vec{\rm u})_{\&} = \partial_{A} u_{\&} \\ \hspace*{6ex} + \;
ig \, [n/2(n+1)]^{1/2} \, u_{\&} \widetilde{C}^{\scriptscriptstyle 0}_{\; A}
\, + g u_{j} \, X^{j}_{\, A} ,
\end{array} \right. $ \hfill \rule{0ex}{4ex}(66b)
\end{flushright} \setcounter{equation}{66}
and
\begin{flushright} $ \left. \begin{array}{l}
(\plaision_{A} \widetilde{\rm v})^{i} = \partial_{A} v^{i} + (i/2) \, g \,
(t_{a})^{i}_{\, j} \, \widetilde{C}^{a}_{\; A} \, v^{j} \\ \hspace*{7ex}
+ \; ig \, [2n(n+1)]^{-1/2} \, \widetilde{C}^{\scriptscriptstyle 0}_{\, A}
\, v^{i} - g X^{i}_{\, A} \, v^{\&} \end{array} \right. $ \hfill (67a) \\
\hspace*{0.75ex} $(\plaision_{A} \widetilde{\rm v})^{\&}
= \partial_{A} v^{\&} - ig \, [n/2(n+1)]^{1/2} \,
\widetilde{C}^{\scriptscriptstyle 0}_{\, A} \, v^{\&},$\rule{0ex}{4ex}
\hfill (67b) \end{flushright} \setcounter{equation}{67}
where $u_{\Theta} = u^{\Theta}$, $v_{\Theta} = v^{\Theta}$,
$X^{i}_{\, A} = X_{iA}$, $\widetilde{C}^{\scriptscriptstyle 0}_{\, A}
= - C^{\scriptscriptstyle 0}_{\, A}$, $\widetilde{C}^{a}_{\, A} =
- \varepsilon^{a}_{b} C^{b}_{\, A}$, and the coefficients
$\varepsilon^{a}_{b}$ are the same as above. Comparing expressions (66) and
(67) with expressions (64) and (63) respectively, we see that at $X_{iA}
\neq 0$ they {\em do not coincide}. Consequently, the interaction with the
fields $X_{iA}$ is not $C$-invariant, and one should observe that in this
case the charge asymmetry is implemented directly in the nonspacetime
degrees of freedom of the fields.

The field strength tensor for the considered gauge fields can be defined
in the standard way. In an arbitrary regular five-vector basis with zero
commutator the components of this tensor are given by the following evident
formula:
\begin{displaymath}
F^{\Theta}_{\; \; \Xi \, AB} = \partial_{A} C^{\Theta}_{\; \Xi B} -
\partial_{B} C^{\Theta}_{\; \Xi A} + C^{\Theta}_{\; \Omega A}
C^{\Omega}_{\; \Xi B} - C^{\Theta}_{\; \Omega B} C^{\Omega}_{\; \Xi A}.
\end{displaymath}
At $\Theta = i$ and $\Xi = j$, with account of equations (56) and (58),
one obtains
\begin{displaymath} \begin{array}{l}
\! F^{i}_{\; j \, AB} = \partial_{A} C^{i}_{\; jB} - \partial_{B}
C^{i}_{\; jA} + C^{i}_{\; kA} C^{k}_{\; jB} - C^{i}_{\; kB} C^{k}_{\; jA} \\
\hspace*{4ex} \rule{0ex}{3ex} = \partial_{A} B^{i}_{\; jB} - \partial_{B}
B^{i}_{\; jA} + B^{i}_{\; kA} B^{k}_{\; jB} - B^{i}_{\; kB} B^{k}_{\; jA},
\end{array} \end{displaymath}
so these components coincide with those of the field strength tensor for
the gauge fields associated with the ordinary nonspacetime vectors that
correspond to the considered $(n+1)$-vectors. In a similar manner one
finds that
\begin{equation} \begin{array}{l}
F^{i}_{\; \& \, AB} = \partial_{A} C^{i}_{\; \& B} + C^{i}_{\; kA}
C^{k}_{\; \& B} \\ \hspace*{10ex} \rule{0ex}{3ex} + \; C^{i}_{\; \& A}
C^{\&}_{\; \& B} - ( A \leftrightarrow B ) = 0
\end{array} \end{equation}
and
\begin{displaymath} \begin{array}{l}
F^{\&}_{\; \& \, AB} = \partial_{A} C^{\&}_{\; \& B} - \partial_{B}
C^{\&}_{\; \& A} + C^{\&}_{\; \& A} C^{\&}_{\; \& B} \\
\hspace*{11ex} \rule{0ex}{3ex} - \; C^{\&}_{\; \& B} C^{\&}_{\; \& A}
= \partial_{A}C^{\&}_{\; \& B} - \partial_{B} C^{\&}_{\; \& A}, \\
F^{\&}_{\; \; j \, AB} = \partial_{A} C^{\&}_{\; jB} + C^{\&}_{\; kA}
C^{k}_{\; jB} \rule{0ex}{3ex} \\ \hspace*{11ex} \rule{0ex}{3ex} +
\; C^{\&}_{\; \& A} C^{\&}_{\; jB} - ( A \leftrightarrow B ).
\end{array} \end{displaymath}
In terms of the fields $X_{jA}$ the latter formula can be rewritten as
\begin{displaymath} \begin{array}{l}
g^{-1} F^{\&}_{\; \; j \, AB} \\ \rule{0ex}{3ex}\hspace*{4ex} =
\partial_{A} X_{jB} - X_{kB} C^{k}_{\; jA} + C^{\&}_{\; \& A} X_{jB}
- \; (A \leftrightarrow B) \\ \rule{0ex}{3ex}\hspace*{4ex} = \partial_{A}
X_{jB} - (ig/2) \, X_{kB} \, (t_{a})^{k}_{\, j} \, C^{a}_{\; A} \\
\rule{0ex}{3ex}\hspace*{7ex} - \; (ig) \, [(n+1)/2n]^{1/2} \,
C^{\scriptscriptstyle 0}_{\, A} \, X_{jB} - \; ( A \leftrightarrow B )
\end{array} \end{displaymath}
or
\begin{equation}
F^{\&}_{\; \; j \, AB}=g \, ({\bf d \widetilde{X}})^{\&}_{\; \, j \, AB}\, ,
\end{equation}
where $\widetilde{\bf X}$ is the five-vector 1-form with components
$(\widetilde{\bf X})^{\&}_{\; \, jA} = X_{jA}$, whose values are elements
of the tensor product of the subspace $W^{\cal E}$ and the subspace
$\widetilde{W}^{\cal Z} \subset \widetilde{W}$ of linear forms on $W$ that
have zero contraction with the $(n+1)$-vectors from $W^{\cal E}$. All this
is in agreement with the fact that with transition from one regular basis in
$\WW$ to another the fields $C^{\&}_{\; jA}$ transform according to formula
(60), i.e.\ according to the anti-fundamental representation with respect
to ${\rm SU}(n)$ and with the charge equal to the difference of the charges
of $u^{\&}$ and $u^{i}$, with respect to ${\rm U}(1)$. In a similar manner,
in terms of the fields $C^{a}_{\, A}$ and $C^{\scriptscriptstyle 0}_{\, A}$,
the formulae for $F^{i}_{\; j \, AB}$ and $F^{\&}_{\; \& \, AB}$ acquire the
form
\begin{equation} \begin{array}{l}
F^{i}_{\; j \, AB} = (ig/2) \, (t_{a})^{i}_{\, j} \, F^{a}_{\; AB} \\
\rule{0ex}{3ex} \hspace*{10ex} + \; (ig) \, [2n(n+1)]^{-1/2} \,
\delta^{i}_{\, j} \, F^{\scriptscriptstyle 0}_{\; AB}
\end{array} \end{equation} \begin{equation}
F^{\&}_{\; \& \, AB} = (-ig) \, [n/2(n+1)]^{1/2} \,
F^{\scriptscriptstyle 0}_{\; AB}, \hspace*{4ex}
\end{equation}
where
\begin{displaymath}
F^{a}_{\; AB} = \partial_{A} C^{a}_{\; B} - \partial_{B} C^{a}_{\; A}
- g f^{a}_{\; bc} C^{b}_{\; A} C^{c}_{\; B}
\end{displaymath} and \begin{displaymath}
F^{\scriptscriptstyle 0}_{\; AB} = \partial_{A}
C^{\scriptscriptstyle 0}_{\; B} -  \partial_{B}
C^{\scriptscriptstyle 0}_{\; A},
\end{displaymath}
and the structure constants $f^{a}_{\; bc}$ are defined as follows:
$[t_{a}, t_{b}] = 2 i \, t_{c} f^{c}_{\; ab}$.

\vspace{3ex} \begin{flushleft}
\bf Acknowledgements
\end{flushleft}
I would like to thank V. D. Laptev for supporting this work. I am grateful
to V. A. Kuzmin for his interest and to V. A. Rubakov for a very helpful
discussion and advice. I am indebted to A. M. Semikhatov of the Lebedev
Physical Institute for a very stimulating and pleasant discussion and to
S. F. Prokushkin of the same institute for consulting me on the Yang-Mills
theories of the de Sitter group. I would also like to thank L. A. Alania,
S. V. Aleshin, and A. A. Irmatov of the Mechanics and Mathematics Department
of the Moscow State University for their help and advice.

\end{document}